# Ce(III) Ions in Hydroxyapatite: Nanoscale Environment Investigation

Sadovnikova, M.A.* (http://orcid.org/0000-0002-5255-9020)[a], Shurtakova, D.V. (http://orcid.org/0000-0003-4765-0724)[a]*, Mamin, G.V. (http://orcid.org/0000-0002-7852-917X)[a], Murzakhanov, F.F. (http://orcid.org/0000-0001-7601-6314)[a], Zobkova, Yu.O.[b], Petrakova, N.V.[b], Komlev, V.S.[b], Gafurov, M.R. (http://orcid.org/0000-0002-2179-2823)[a].

[a] Institute of Physics, Kazan Federal University, Kremlyovskaya 18, Kazan 420008, Russia

[b] A.A. Baikov Institute of Metallurgy and Materials Science RAS, Leninsky Avenue 49, 119334 Moscow, Russia

*e-mail: margaritaasadov@gmail.com, darja-shurtakva@mail.ru

**Abstract**

This paper presents a comprehensive study of cerium-doped hydroxyapatite (Ce-HAp), a material of interest for biomedical applications due to the good biocompatibility of hydroxyapatite and the antioxidant activity of cerium ions. We employ advanced electron paramagnetic resonance (EPR) techniques, including continuous wave (CW) and pulsed X-band experiments, electron spin echo envelope modulation (ESEEM), and electron-electron double resonance (ELDOR)-detected NMR (EDNMR) to investigate the local coordination and electron environment of cerium ions in the hydroxyapatite matrix. The experimental results are complemented by g-tensor calculation, which allows us to interpret the EPR spectra and identify the types of paramagnetic centers. Our results show that during the synthesis of hydroxyapatite powder by the chemical precipitation method using nitrates, cerium ions enter the structure mainly in the trivalent state and replace calcium ions in two nonequivalent positions. In addition to cerium ions, nitrate radicals are found in the HAp structure. Heat treatment reduces the amount of nitrate radicals and increases crystallinity. This work expands the understanding of the role of cerium in calcium phosphates and provides a methodological basis for the characterization of doped bioceramics using various EPR approaches.

**Introduction**

In recent years, hydroxyapatite (HAp) has attracted considerable attention due to its unique combination of features, namely biocompatibility, high bioactivity, no toxicity, tunable properties and versatility in various applications [1]. Its similarity in both chemical composition and structure to the phosphate systems of bone has led to its widespread use in bone grafting and implantation, as well as in dental materials. The presence of porosity provides high binding capacity for various pharmacological substances such as antibiotics, hormones, enzymes, steroids, etc [2]. This opens up the potential for using synthetic HAp to deliver pharmacological substances in many clinical applications with the possibility of sustained release for the treatment of osteomyelitis, osteoporosis, bone cancer. Nevertheless, the properties of HAp may be limited to meet the growing challenges of modern medical technology. To expand the application range of HAp and improve its functionality, doping with various ions is actively explored. HAp is widely known as a matrix for introducing new elements due to its flexible structure, which allows ionic substitution of $Ca^{2+}$, $PO_4^{3-}$, and $OH^-$ ions. These substitutions can include monovalent cations ($Na^+$, $K^+$, etc.) to improve the body's homeostasis and bone tissue demineralization processes; divalent cations ($Sr^{2+}$, $Pb^{2+}$, $Ba^{2+}$, $Zn^{2+}$, $Mn^{2+}$, $Cd^{2+}$, $Mg^{2+}$, etc.) to enhance osteogenesis processes, antibacterial and antifungal activity; trivalent ($Cr^{3+}$, $Al^{3+}$, $Fe^{3+}$, rare earth ions, etc.), tetravalent ($Ti^{4+}$, $Th^{4+}$, $U^{4+}$) and even hexavalent cations ($U^{6+}$) to provide magnetic, catalytic, antibacterial and luminescent properties [3-5].

Rare earth elements (REEs) significantly enhance the biomedical potential of HAp due to their unique physicochemical properties (luminescence, magnetic properties, stimulation of osteogenesis, contrast agent for bioimaging). Diagnostics and imaging play an important key role in primary detection, screening and image-guided smart nanomedicine for medical solutions. REEs, including cerium, are promising imaging agents in the field of bioimaging for the diagnosis and treatment of damaged organs and tissues, due to their ability to luminesce in a wide range of electromagnetic radiation. In the literature, the co-occurrence of cerium and calcium is reported in natural compounds, namely in apatites [6]. Cerium has been found to be able to participate in calcium-dependent reactions, which can also stimulate metabolism and accumulate in small amounts in bone tissue [7]. As early as the end of the 19th century, the bacteriostatic effect of cerium was studied, leading to its use as a local antiseptic [8]. Many sources point to the antitumour activity of cerium (IV) oxide and its potential as an antioxidant and radioprotector in the treatment of oncological diseases [9].

Cerium ions are also characterised by luminescence in the UV region of light, caused by 4f-f or 4f-5d electron transitions [10]. The transition of an electron between energy levels in stable $Ce^{3+}$ is accompanied by a fast and bright luminescence, whereas in $Ce^{4+}$ the luminescence is insignificant [11]. In compounds, cerium is present in a mixed form, both $Ce^{3+}$ and $Ce^{4+}$, due to the environment of external influence on the material [12]. The intensity of the luminescence of cerium-containing material depends on the quantitative content of $Ce^{3+}$ ions, the presence of impurities of different nature and the characteristics of the material particles [12].

Cerium ($Ce^{3+}$) ions exhibit luminescent properties, allowing $Ce^{3+}$ to be used in optical imaging, such as fluorescence microscopy and spectroscopy. Cerium nanoparticles can be functionalized for targeted delivery to specific cells or tissues, increasing the specificity of imaging. The unique features of Ce ions such as antioxidant activity, bioresorbable modulation and potential radioprotection make them a promising dopant for HAp [13]. The incorporation of Ce into HAp is expected to improve both biocompatibility and bone tissue regeneration, thus increasing the attractiveness of these composites for implant applications. This study illustrates the successful fabrication of luminescent cerium-doped hydroxyapatite (Ce-HAp).

The objective of the present study is to establish the influence of synthesis and processing methodologies on the content of foreign impurities, magnetic properties, and structural modifications in Ce-HAp powders using various electron paramagnetic resonance (EPR) techniques. To obtain more specific structural information about the role of these dopants at the atomic level, suitable spectroscopic techniques are required. EPR is a powerful tool for studying the structural and electronic properties of materials, including REE doped HAp. EPR allows the detailed study of the interactions between the alloying elements and the HAp matrix and their effect on the magnetic and optical properties. This study is important for understanding the mechanisms underlying the improvement of the biocompatibility and functionality of HAp, which in turn may contribute to their wider application in clinical practice. The present work is devoted to the characterization of Ce-HAp, with an emphasis on the relationship between the composition, structure and properties of the obtained materials, as well as on the prospects for their use in biomedicine.

**Materials**

**Powder synthesis**

Ce-HAp were synthesized by chemical precipitation according to the reaction (1). A 0.1 M solution of cerium (III) nitrate was added to a 0.5 M solution of calcium nitrate, and a 0.5 M solution of ammonium hydrogen phosphate was added dropwise to the resulting mixture. A 25% solution of ammonia was added to maintain the required pH of the medium (11±0.5) [4].

$$(10 - x)Ca(NO_3)_2 + (2x/3)Ce(NO_3)_3 + 6(NH_4)_2HPO_4 + 8NH_4OH \rightarrow$$
$$\rightarrow Ca_{(10-x)}Ce_{(2x/3)}(PO_4)_6(OH)_2 + 6H_2O + 20NH_4NO_3, \qquad (1)$$

where x is a degree of substitution, $[Ca^{2+}]$ is an atomic content of $Ca^{2+}$ in HAp, $[Ce^{3+}]$ is an atomic content of $Ce^{3+}$ in Ce-HAp; x = ($[Ca^{2+}] \times [Ce^{3+}]$)/100; $[Ce^{3+}]$ = 0 and 0.5 for $[Ca^{2+}]$.

The amount of Ce ([Ce]) was calculated of 0.5 in respect to Ca atomic content in HAp ([Ca]): thus, for HAp [Ca]=10, the amount of Ce was taken of 0.5 for [Ca], therefore [Ce] = (10 × 0.5)/100. For this, the 0.5 M aqueous solution of ammonium hydrophosphate ($(NH_4)_2HPO_4$) was dropwise added to the mix of 0.5 M solution of calcium nitrate ($Ca(NO_3)_2$) and 0.1 M solution of cerium nitrate ($Ce(NO_3)_2$) in appropriate amount under constant stirring. The pH level was adjusted at 11.0 ± 0.5 by adding the ammonium hydroxide solution ($NH_4OH$). The resulted suspensions were kept for aging for 24 h under room conditions. The precipitate was filtered on a Buchner funnel, dried at 70 °C, ground in a corundum mortar, and sifted through a sieve with a mesh size of 100 μm.

It is known that it is possible to prevent the oxidation processes of Ce(III) to Ce(IV) under the conditions of a reducing or protective environment. To achieve a reducing atmosphere, heat treatment and hot pressing methods have been used. As-dried powders were heat treated under different conditions: (i) the heat treating at 1300°C in air atmosphere (Ce-HAp-air) in a chamber furnace with SiC heaters and (ii) hot pressing (Ce-HAp-HP) in a hot press furnace (Thermal technology Inc., model HP 250 3560-20) in a carbon mold in an argon atmosphere at a pressure of 30 MPa and temperature of 1100 °C to achieve the reducing atmosphere.

**Crystal structure**

Naturally occurring HAp has a hexagonal in structure (space group P63/m, two formula units per unit cell) with lattice parameters $a = b = 0.942$ nm and $c = 0.687$ nm, and the chemical formula of a unit cell is $Ca_{10}(PO_4)_6(OH)_2$.

The atomic structure of HAp is shown in Fig. 1. The Ca atoms reside in two positions: 6 atoms per unit cell are in position Ca(II) and 4 atoms are in position Ca(I). Ca(I) is located on the threefold axis and is coordinated by 9 oxygens of phosphate groups. The Ca(II) atoms are bonded to seven oxygen atoms in the first coordination sphere: six of them are related to the four orthophosphate groups, and the seventh is related to the hydroxyl group.

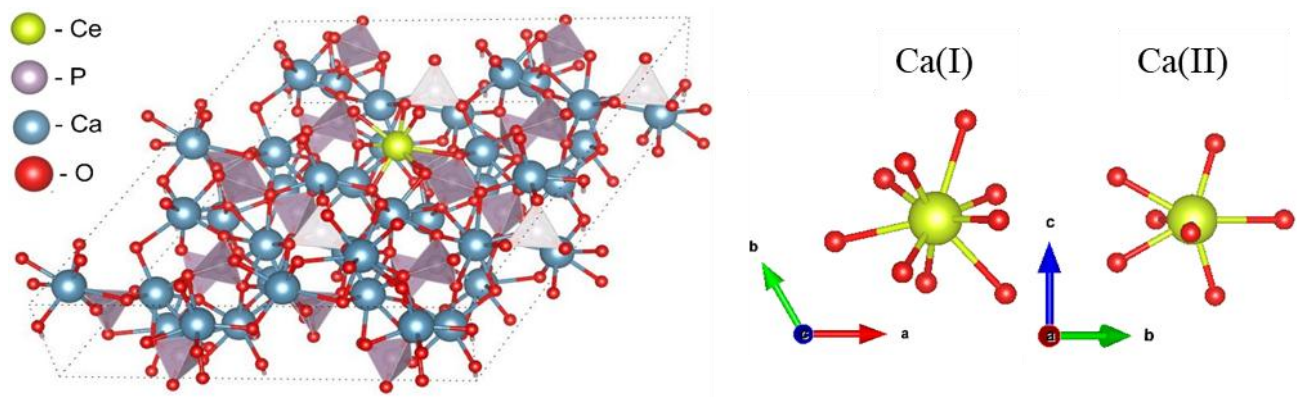

Fig. 1 Crystal lattice of HAp doped with Ce ions

Cationic substitutions in the HAp structure allow the modification of material properties. The fundamental aspects in the study of the substituted forms of HAp are the localization of the impurity ions in the structure, their distribution by the coordination positions of the calcium ion Ca(I)/Ca(II).

The difference in valence between the calcium ion and the REEs leads to the need for charge balancing during the substitution. The substitution of a $Ca^{2+}$ ion by a $Ce^{3+}$ ion creates an excess positive charge in the lattice. Charge compensation is one of the main conditions of the process and can be achieved in three main ways. In the first case, the substitution of a higher-charged ion for a lower-charged ion is accompanied by the formation of a vacancy. In this case, the structure of the material becomes defective. In the second case, valence compensation is accompanied by the introduction of a single-charged alkali metal cation into the interstitial spaces. The third type of heterovalent substitution involves charge compensation by atoms of different valence. Considering that the incorporation of low concentrations of cerium ions into the HAp structure does not induce significant structural changes, the most probable compensation mechanism in this case is the formation of a proton vacancy in the $OH^-$ group located in the anion channel, as such a defect position has minimal impact on the unit cell density and atomic packing [5].

**Methods**

**X-ray Diffraction**

The X-ray Diffraction (XRD) patterns were collected on XRD-6000 powder diffractometer (Bragg-Brentano geometry, Scintillator detector, CuKα radiation, λ = 1.5418Å, Shimadzu, Japan). The XRD data were recorded at room temperature in the 2θ range between 10° and 60°, with intervals of 0.02°, and a counter speed of 2°/min. The element content of cerium for the Ce-HAp powders was determined using the XFS Spectroscan MAKC-GVM spectrometer (St. Petersburg, Russia).

**Electron Paramagnetic Resonance**

EPR measurements were performed in the continuous wave (CW) and pulsed modes in X-band ($v_{mw}$ = 9.6 GHz) on a Bruker Elexsys E580 spectrometer (Billerica, MA, USA). Field-swept electron spin-echo (ESE) spectra were recorded with the standard Hahn pulse sequence π/2 − τ − π with a π/2 pulse duration of 16 ns, and a minimal time delay τ = 240 ns. $T_2$ was studied by tracking the primary ESE amplitude with the same π/2 − π pulse durations while varying τ with a minimal possible step of 4 ns. $T_1$ was extracted from an inversion recovery experiment by applying the π − $T_{delay}$ − π/2 − τ − π pulse sequence, where both the π pulse duration and τ are fixed while Tdelay is varied. Electron-nuclear interactions were analyzed using a three-pulse Electron Spin Echo Envelope Modulation (ESEEM) sequence (π/2–τ–π/2–T–π/2–ESE), with distances T varying from 180 ns to the required values (8–64 μs). The ESEEM experiment was carried out at T = 10 K. The electron-electron double resonance (ELDOR)-detected NMR (EDNMR) method for electron-nuclear research is based on the double electron-electron resonance approach. To study double electron-electron transitions, an additional ELDOR module E580-400U and a dielectric ring resonator ER 4118X-MD5 were used. In this case, the main source of microwave radiation is the observation frequency ($v_{mw1}$) of a fixed value ($v_{mw}$ = 9.59 GHz). The E580-400U module is required to generate and change the second independent microwave frequency ($v_{mw2}$) in the range from 9.3 to 10 GHz. The ER 4118X-MD5 resonator, due to its adjustable quality factor, allows optimizing the resonance frequency band in accordance with the experimental requirements.

To investigate the radiation-induced paramagnetic species, X-ray irradiation of the materials was performed using a URS-55 source (U = 50 kV, I = 15 mA, W-anticathode) at T = 297 K for 1 h with a calculated dose of 15 kGy.

**Calculation of g-tensor**

The Hamiltonian for the $Ce^{3+}$ impurity ion in hydroxyapatite under an external magnetic field can be expressed as follows [14]:

$$H = H_{FI} + H_{SO} + H_{CF} + H_Z, \qquad (2)$$

where $H_{FI}$ is the free ion Hamiltonian, encompassing the kinetic energy of all electrons and the Coulomb interaction (not considered in this work, as it does not lead to energy level splitting); $H_{SO}$ is the spin-orbit interaction Hamiltonian; $H_{CF}$ is the crystal field Hamiltonian; and $H_Z$ is the Zeeman interaction Hamiltonian.

The spin-orbit interaction Hamiltonian for the $Ce^{3+}$ impurity ion can be written as:

$$H_{SO} = \xi \mathbf{SL}, \qquad (3)$$

where $\xi$ is the spin-orbit coupling constant, $\mathbf{S}$ and $\mathbf{L}$ are the spin and orbital angular momentum operators, respectively. The $Ce^{3+}$ ion has a $4f^1$ electronic configuration with a $^2F$ ground term. Due to spin-orbit interaction, the $^2F$ term splits into $^2F_{5/2}$ and $^2F_{7/2}$, where the lower-energy state is $^2F_{5/2}$, and the splitting energy corresponds to 2200 cm$^{-1}$ [14].

The crystal field Hamiltonian can be written as [15]:

$$H_{CF} = \sum_k \sum_{q=-k}^{k} B_q^{(k)} C_q^{(k)}, \qquad (4)$$

$$B_q^{(k)} = \sum_j a^{(k)}(R_j)(-1)^q C_{-q}^{(k)}(\theta_j, \varphi_j), \qquad (5)$$

where $B_k^q$ are the crystal field parameters, $C_{-q}^{(k)}(\theta_j, \varphi_j) = \sqrt{4\pi/(2k+1)} Y_q^k$ are the spherical tensors, and $a^{(k)}(R_j)$ include contributions from lattice ions, exchange and covalent contributions, and the contribution from the overlap of the electron orbits of the cerium ion and the outer shells of nearby oxygen ions. These contributions are discussed in detail in [15].

Since the $Ce^{3+}$ ion's ground state is $^2F_{5/2}$ with J = 5/2, the non-zero crystal field parameters will correspond to q = 2, 4. In hydroxyapatite, the $Ce^{3+}$ ion substitutes for the $Ca^{2+}$ ion. The $Ca^{2+}$ ion occupies two non-equivalent lattice sites: Ca(I) with $C_3$ symmetry and Ca(II) with $C_{1h}$ symmetry.

By diagonalizing the crystal field matrix, we obtain the energy eigenvalues and corresponding wave functions of the $^2F_{5/2}$ ground state split into three levels, each representing a Kramers doublet. The wave functions of the Kramers doublet are related by the time-reversal operator $\hat{\theta}$ [16]:

$$|\psi\rangle = \sum c_{JM} |J, M\rangle, \qquad (6)$$

$$|\psi'\rangle = \hat{\theta}|\psi\rangle = \sum c_{JM}^* (-1)^{J-M} |J, M\rangle,$$

Let us consider the electron Zeeman energy within the main *J*-multiplet [16]:

$$H_Z = \mu_B g_J \mathbf{B} \cdot \mathbf{J} = \mu_B g_J \sum_\alpha B_\alpha J_\alpha = \mu_B \sum_{\alpha,\beta} g_{\alpha\beta} B_\alpha S_\beta, \qquad (7)$$

where

$$g_{\alpha z} = 2g_J\langle\psi|J_\alpha|\psi\rangle, \quad g_{\alpha x} = 2g_J Re\langle\psi|J_\alpha|\psi'\rangle, \quad g_{\alpha x} = -2g_J Im\langle\psi|J_\alpha|\psi'\rangle, \quad (8)$$

$g_J$ is the Lande factor, $g_J = 6/7$ for the $Ce^{3+}$ ion [14].

Note that the values of the g-tensor in this case depend on the definition of the wave functions $|\psi\rangle$ and $|\psi'\rangle$, while the value defined as $\mathbf{G}=\mathbf{g}\mathbf{g}^T$ is an invariant of the energy level splitting. Therefore, the principal values of the g-factor ($g_x$, $g_y$, $g_z$) can be found by diagonalizing the $\mathbf{G}$-tensor and calculating the corresponding eigenvalues [17].

**Results and Discussion**

**XRD Analysis**

According to XRD data (Fig. 2), HAp is the main crystalline phase in the synthesized samples (JCPDS card [09-0342]).

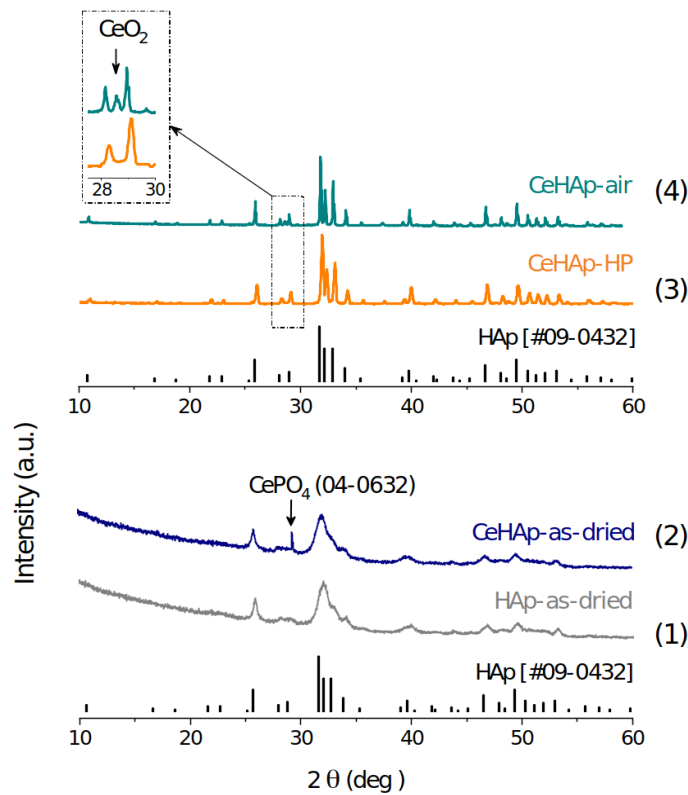

Fig. 2 XRD patterns for undoped HAp and Ce-HAp: (1) – pattern for undoped as-dried HAp quite consistent with the HAp card [09-0432], (2) – pattern for as-dried Ce-HAp indicates the presence of CePO4 [04-0632], (3) – pattern for Ce-HAp obtained by hot pressing in reducing atmosphere fully correspond to HAp card without any impurity; (4) – pattern for Ce-HAp obtained by heat treating in air atmosphere contains the $CeO_2$ impurity

For the Ce-HAp (2) sample, the peak detected at 29.3 was assigned to cerium phosphate $CePO_4$, which was not present in the scheme for undoped HAp. The X-ray diffraction patterns for Ce-HAp-air (4) and Ce-HAp-HP (3) were consistent with well crystallized HAp. The pattern for Ce-Hap (4) contained a weak peak at 28.5 assigned to cerium oxide $CeO_2$.

**EPR Analysis**

It is an established fact that HAp powders in their own crystal lattices do not contain ions with non-zero magnetic moments. Consequently, EPR signals from samples without impurities will be absent. The presence of resonance absorptions from nominally "pure" samples is indicative of an uncontrolled impurity in the sample structure, resulting from the reagents used or the formation of various side phases during synthesis.

**Continuous wave EPR** spectrum recorded under room conditions for a Ce-HAp sample shown in Fig. 3. $Ce^{3+}$ is a Kramer's ion and its magnetic properties arise from a partially filled 4f shell. $Ce^{3+}$ in the ground state has an electron spin $S = ½$ and an orbital angular momentum $L = 3$, which leads to very short relaxation times due to spin-orbit and spin-lattice interactions. Accordingly, when the EPR spectrum is recorded at room temperature, the absorption signal related to the rare earth cerium is broadened, so that there is no resonance signal within the sensitivity of the spectrometer.

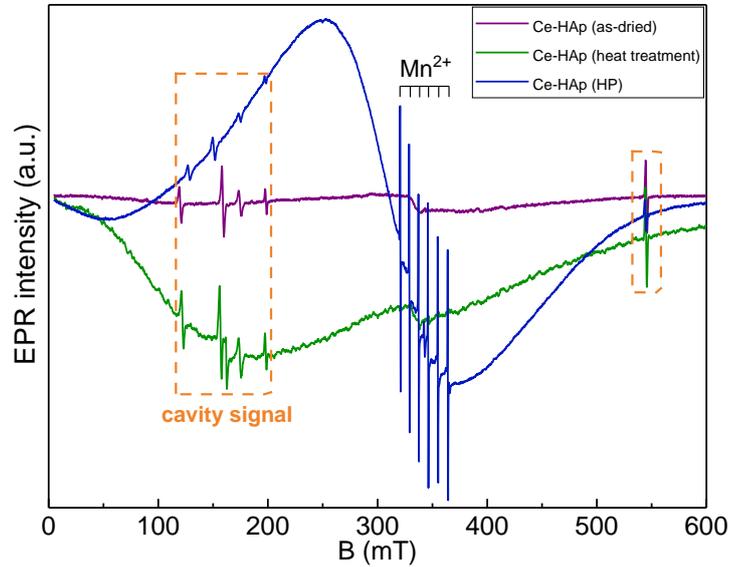

Fig. 3 EPR spectra recorded in CW mode in the X-band at room temperature for Ce-HAp (as-dried); Ce-HAp (heat treatment at 1300º); Ce-HAp (HP 1100ºC) Signals marked with an orange frame are related to the spectrometer cavity

From Fig. 3 it can be concluded that impurities (by-products of synthesis) are absent for the Ce-Hap (as-dried) and the sample after heat treatment at 1300ºC in air. It is also noted from the EPR spectrum that the powders subjected to hot pressing (Fig. 3, blue curve) contain various impurities, namely 6 narrow lines characteristic of manganese ions $Mn^{2+}$ (Fig. 3, blue curve). The presence of splitting at the g-factor $\approx 2$ is caused by the hyperfine interaction between the electron $S = 5/2$ and nuclear $I = 5/2$ magnetic moments with a value of 93 G. However, these narrow-resolved lines with peak widths of about 2-3 G each and the absence of zero splitting indicate that $Mn^{2+}$ ions are not incorporated into the HAp crystal lattice. The appearance of a broad line (Fig. 3, blue) may also indicate the presence of iron oxide groups on the surface of the HAp nanoparticles.

**Echo-detected EPR** spectra were recorded using the technique of spin echo and magnetic field sweep. The results for Ce-HAp were obtained by detecting the integral intensity of the electron spin echo (ESE) depending on the magnetic field strength. For all further experiments the sample temperature was lowered to $T = 10$ K in order to increase the relaxation time and to improve the resolution. In fact, no resonance lines were detected above 20 K, due to the rapid relaxation time $T_1$ which broadens the line width (line width $\propto T_1^{-1}$), as observed for most lanthanides with the exception of S-state ions with a half-field subshell $4f^7$.

EPR spectra for Ce-HAp after heat treated in air 1300º were obtained in pulsed mode using the Hahn sequence (Fig. 4, blue insert). The line shape of this broad signal arises from the $^2F_{5/2}$ ground state of $Ce^{3+}$ ions and is likely influenced by both the anisotropic zero-field splitting and the anisotropic g-tensor. Broad and poorly resolved line shapes of this kind have been observed for various $Ce^{3+}$-doped glasses [18].

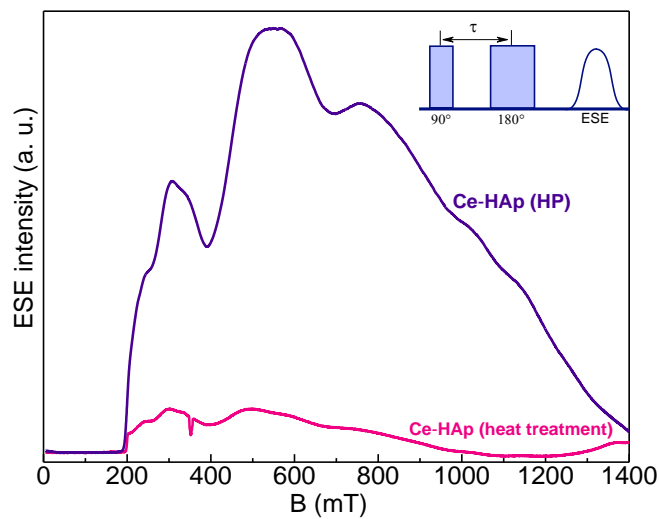

Fig. 4 Comparative analysis ESE-EPR spectra for Ce-Hap after heat treatment and HP

For Ce-HAp subjected after hot press 1100⁰ the EPR spectrum is shown in Fig. 4 (purple line). In the ESE spectrum, a broad signal is observed.

For clarity, the ESE spectra for two different treatments after synthesis are plotted on one graph (Fig. 4). From the obvious difference in signal intensity, it can be concluded that after HP, cerium is more in the trivalent state, whereas after heat treatment in air, cerium ions are partially transformed into the tetravalent state. The Ce-Hap as-dried powder has no absorption signal, indicating the presence of 4+ valence cerium.

Table 1 shows the relaxation times (spin-spin $T_2$ and spin-lattice $T_1$). The samples obtained after the hot press have shorter relaxation times, which further indicates the presence of cerium in the trivalent state. Also, since $T_2$ relaxation is much shorter for samples after a hot press, we can conclude that more cerium remains after a HP than after heat treatment in air.

Table 1. Relaxation times for Ce-HAp at T = 10 K:

|  | $T_1$, µs | $T_2$, µs | B, mT |
| --- | --- | --- | --- |
| Ce-HAp (heat treatment 1300⁰) | 127 | 2.4 | 300 |
| Ce-HAp (HP) | 57 | 0.4 | 300 |

**Calculation of g-tensor**

To calculate the g-tensor, a program was written in Python. The program performs the following tasks:

1. Calculation of crystal field parameters by eq.5.

2. Generates the crystal field Hamiltonian matrix by eq.4

3. Matrix Diagonalization: Diagonalizes the crystal field Hamiltonian matrix to obtain the eigenenergies and eigenfunctions of the split energy levels (e.g., Kramers doublets).

4. Calculation the effective *g*-tensor using the derived wave functions and the Zeeman interaction formalism by eq.8

The input data for the computational model consist of the coordinates of the nearest oxygen ligands, extracted from the optimized crystal structure. The geometry optimization and electronic structure calculations were performed using the Quantum Espresso software package, which employs density functional theory (DFT) with the generalized gradient approximation (GGA) for the exchange-correlation functional, specifically the Perdew–Burke–Ernzerhof (PBE) parametrization. Kinetic energy cutoff values of 70 Ry for the plane-wave basis set (wavefunctions) and 300 Ry for the charge density were applied. Calculations were conducted using a 2 × 2 × 1 supercell in the P6₃/m space group, containing 44 atoms in the primitive unit cell (176 atoms in the supercell). Geometry optimization was carried out under the assumption of low paramagnetic impurity concentration; only atomic positions were relaxed, while lattice parameters remained fixed to preserve the crystallographic framework.

In the case of an impurity in the Ca(I) site, the ion configuration changed insignificantly. However, the distances to the nearest oxygen atoms were altered. Specifically, the distances to the three closest oxygen atoms increased, while those to the remaining six oxygen atoms decreased. In contrast, substituting $Ce^{3+}$ into the Ca(II) site induced more pronounced structural reorganization: the cerium ion shifted closer to the hydroxyl channel, reducing the distance between $Ce^{3+}$ and the hydroxyl oxygen from 2.388 Å to 2.187 Å. Notably, for both substitution sites (Ca(I) and Ca(II)), the distance between the hydroxyl channel oxygen and the nearest calcium ion decreased from 2.388 Å to a range of 2.167–2.346 Å. The optimized total energies of the two systems were nearly identical (−6791.66 Ry for the Ca(I)-substituted cell and −6791.71 Ry for the Ca(II)-substituted cell), suggesting comparable thermodynamic stability and thus equal probability of $Ce^{3+}$ occupying either crystallographic site. During geometry optimization, the atomic positions underwent slight adjustments, leading to a reduction in the local symmetry of the Ca(I) site. As a result, non-zero crystal field parameters are expected to emerge for the Ca(I) position, reflecting the lowered symmetry, as well as for the Ca(II) position.

The calculated crystal field parameters include contributions from lattice ions $a_{pc}^{(k)}(R_j)$ (point charge model), the exchange interaction $a_{ex}^{(k)}(R_j)$, and the overlap contribution $a_{ec}^{(k)}(R_j)$ arising from the overlap between the electron orbitals of the $Ce^{3+}$ ion and the outer shells of neighboring oxygen ions, following the approach outlined in [15]. In the calculation of the exchange contribution, the parameter of model $G$ was set to $G=9$ for the Ca(I) site and $G=3$ for the Ca(II) site. The computed values of the contributions for $Ce^{3+}$ ions are summarized in Tables 2 and 3. The parameters $a_{pc}^{(k)}(R_j)$ and $a_{ec}^{(k)}(R_j)$ are scaled by the screening factor $(1-\sigma_k)$, where $\sigma_2 = 0.722$ and $\sigma_4 \approx 0$ [19, 20].

Table 2. Calculated crystal field parameters for the Ca(I) position

| cm$^{-1}$ | $a_{pc}^{(k)}(R_j)(1-\sigma_k)$ | $a_{ec}^{(k)}(R_j)(1-\sigma_k)$ | $a_{ex}^{(k)}(R_j)$ | Total |
|---|---|---|---|---|
| $B_2^0$ | -1089 | 717 | -165 | -538 |
| $B_2^1$ | -35-64i | 24+44i | -3-7i | -14-27i |
| $B_2^{-1}$ | 35-64i | -24+44i | 3-7i | 14-27i |
| $B_2^2$ | -19+64i | 13-41i | -1+12i | -6+35i |
| $B_2^{-2}$ | -19-64i | 13+41i | -1-12i | -6-35i |
| $B_4^0$ | -1334 | 1351 | -693 | -676 |
| $B_4^1$ | -39+2i | 37-2i | -25+1i | -27+1i |
| $B_4^{-1}$ | 39+2i | -37-2i | 25+1i | 27+1i |
| $B_4^2$ | 31-88i | -33+97i | 15-32i | 13-23i |
| $B_4^{-2}$ | 31+88i | -33-97i | 15+32i | 13+23i |
| $B_4^3$ | -753-472i | 719+468i | -484-257i | -518-261i |
| $B_4^{-3}$ | 753-472i | -719+468i | 484-257i | 518-261i |
| $B_4^4$ | 8-14i | -6+16i | 10-3i | 13-1i |
| $B_4^{-4}$ | 8+14i | -6-16i | 10+3i | 13+1i |

Table 3. Calculated crystal field parameters for the Ca(II) position

| cm$^{-1}$ | $a_{pc}^{(k)}(R_j)(1-\sigma_k)$ | $a_{ec}^{(k)}(R_j)(1-\sigma_k)$ | $a_{ex}^{(k)}(R_j)$ | Total |
|---|---|---|---|---|
| $B_2^0$ | 145 | -60 | 42 | 127 |
| $B_2^1$ | 158-210i | -130+139i | 3-10i | 31-81i |
| $B_2^{-1}$ | -158-210i | 130+139i | -3-10i | -31-81i |
| $B_2^2$ | -160-507i | 349+302i | 73-47i | 263-252i |
| $B_2^{-2}$ | -160+507i | 349-302i | 73+47i | 263+252i |
| $B_4^0$ | 997 | -1083 | 172 | 87 |
| $B_4^1$ | 315-37i | -395+37i | 33-4i | -46-4i |
| $B_4^{-1}$ | -315-37i | 395+37i | -33-4i | 46-4i |
| $B_4^2$ | -584+535i | 683-531i | -76+94i | 21+98i |
| $B_4^{-2}$ | -584-535i | 683+531i | -76-94i | 21-98i |
| $B_4^3$ | -343-34i | 403+64i | -46+8i | 13+38i |
| $B_4^{-3}$ | 343-34i | -403+64i | 46+8i | -13+38i |
| $B_4^4$ | 971-348i | -1105+198i | 147-142i | 13-291i |
| $B_4^{-4}$ | 971+348i | -1105-198i | 147+142i | 13+291i |

After diagonalizing the crystal field matrix, the $^2F_{5/2}$ ground state splits into three Kramers doublets with the following energies:

Ca(I): $\varepsilon_1 = 0$ cm$^{-1}$, $\varepsilon_2 = 334$ cm$^{-1}$, $\varepsilon_3 = 401$ cm$^{-1}$,

Ca(II): $\varepsilon_1 = 0$ cm$^{-1}$, $\varepsilon_2 = 97$ cm$^{-1}$, $\varepsilon_3 = 284$ cm$^{-1}$.

Wave functions of the lower doublet:

$$\psi = a \left|+\frac{5}{2}\right> + b \left|+\frac{3}{2}\right> + c \left|+\frac{1}{2}\right> + d \left|-\frac{1}{2}\right> + e \left|-\frac{3}{2}\right> + f \left|-\frac{5}{2}\right>,$$

the coefficients are given in Table 4.

Table 4. Coefficients before the wave functions for the lower doublet

| | Ca(I) | Ca(II) |
|---|---|---|
| a | 0,89 | -0,20 |
| b | 0,01-0,03i | -0,08+0,60i |
| c | 0 | -0,02-0,32i |
| d | 0,41-0,21i | 0,51-0,25i |
| e | 0 | -0,12+0,38i |
| f | 0 | -0,03 |

The corresponding wave functions $\psi'$ are found using eq.6. The g-tensor calculated by eq.8 present in table 5. The calculated values of the g-tensor were used as initial values to describe the experimental spectrum. The result of the description is shown in Fig. 5, the values are given in Table 4.

Table 5. The g-tensor calculated by eq.8 and g-tensor obtained by describing experimental spectra ($R_{Ca(I)}$ and $R_{Ca(II)}$ in Fig. 5)

| | | $g_{xx}$ | $g_{yy}$ | $g_{zz}$ |
|---|---|---|---|---|
| Ca(I) | calc. | 0.48 | 0.60 | 3.27 |
| | $R_{Ca(I)}$ | 0.51 | 0.85 | 3.29 |
| Ca(II) | calc. | 0.49 | 1.39 | 2.70 |
| | $R_{Ca(II)}$ | 0.56 | 1.25 | 2.93 |

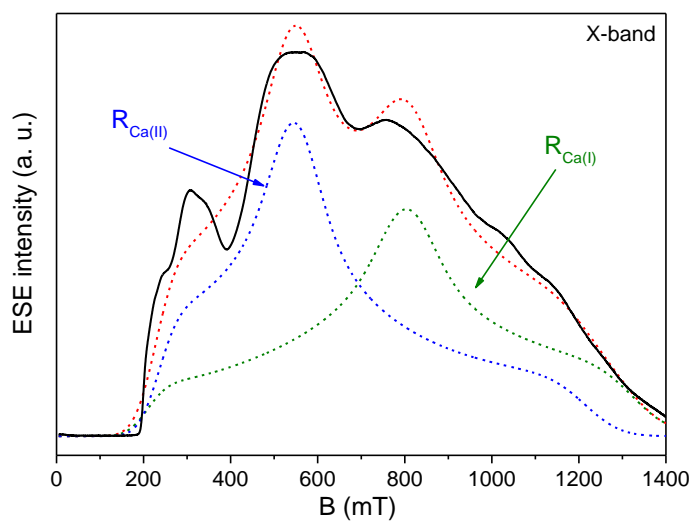

Fig. 5 Comparative analysis ESE-EPR spectra for Ce-Hap after heat treatment and HP

Analysis of the spectrum presented in Figure 5 demonstrates that the experimental absorption spectrum can be attributed to $Ce^{3+}$ ions in two nonequivalent crystallographic positions. The observed discrepancies between the theoretically calculated values of the $g$-tensor and those obtained from the approximation of experimental data are likely caused by a stronger violation of local symmetry in the crystal lattice. Specifically, for the Ca(I) center (point symmetry group $C_3$), the experimentally determined components of the $g$-tensor ($g_x = 0.51$ and $g_y = 0.85$) indicate a reduction in local symmetry. Such deviation from the original symmetry may result from crystal field distortions caused by lattice defects arising from the substitution of calcium ions by cerium ions, or interactions with neighboring ions not considered in this work (e.g., electron-phonon interactions leading to symmetry reduction due to vibrations of neighboring atoms).

**ESEEM**

The electron-nuclear environment was analyzed using the ESEEM method (Electron Spin Echo Envelope Modulation), which allows to detect a signal from magnetic nuclei of a weak interaction. Modulation occurs when the Ce impurity center is coupled to the surrounding nuclei through an anisotropic dipole-dipole interaction. Fig. 6 (a) shows the decay curves of the transverse electron magnetization of cerium ions for two different magnetic fields. The experiments were carried out at 10 K to increase the relaxation times and, as a consequence, to improve the resolution of the ESEEM method. In these curves, in the case of a fixed magnetic field of $B_0 = 304$ mT and 352 mT, intense nuclear modulations can be observed due to the presence of forbidden transitions in the nuclear spin due to the anisotropic dipole-dipole contribution to the hyperfine interaction. To establish the nature of the nearest magnetic core causing the modulations, it was necessary to perform a Fourier transform, the frequency spectrum of which is shown in Fig. 6 (b). The obtained ESEEM results in the time range were processed by Origin Pro 2017, by subtracting the exponential curve to obtain the nuclear modulations.

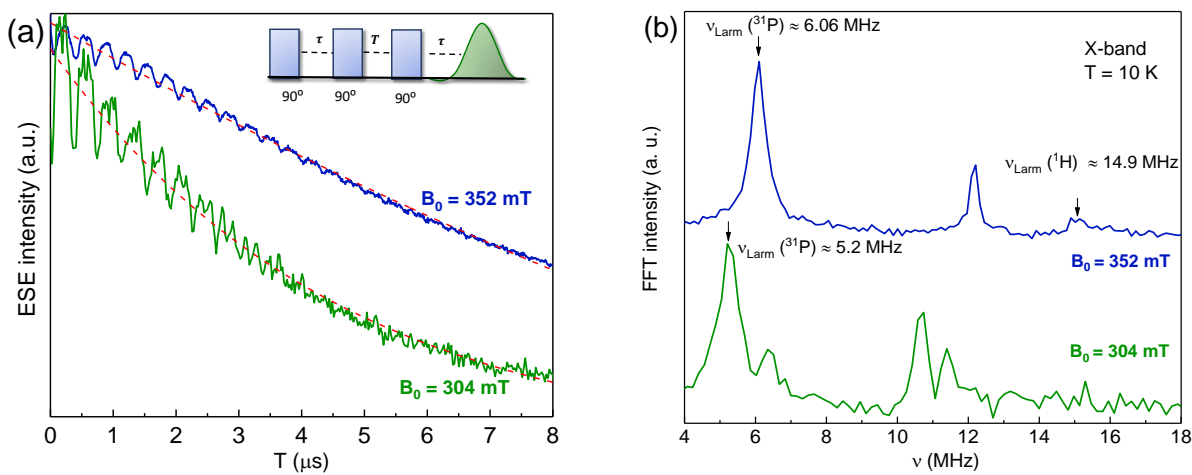

Fig. 6 (a) ESEEM for two different values $B_0$; (b) spectrum of nuclear transitions in the frequency range

The ESEEM spectrum shows signals in the region of 6 MHz and 5.2 MHz, depending on the fixed magnetic field. Taking into account the gyromagnetic ratio of the magnetic nuclei and the magnitude of the magnetic induction, it can be established that the observed signals correspond to the Larmor frequencies of phosphorus nuclei $^{31}P$ with spin I = ½.

The hyperfine interaction was calculated, which corresponds to the distance between the cerium ions and phosphorus nuclei according to eq. 9. The calculation results (Tab. 6) correspond to the distance between the cerium ions and phosphorus nuclei of the order of 3-4 Å, which corresponds to the interatomic distances between the calcium ions and phosphorus groups $PO_4$, then it can be argued about the localization of Ce ions in the HAp structure, and not on the surface of nanocrystals or in another side phase. The observed signal in the region of 14.9 MHz at $B_0$ = 352 mT corresponds to the Larmor frequency of hydrogen nuclei $^1H$ with spin I = ½. Hydrogen is not a structural element of HAp, accordingly, it can be assumed that it either appears as a charge compensator or is localized in the interstitial site of the crystal lattice.

The magnitude of this interaction is related to the distance through the anisotropic dipole-dipole approximation and is expressed by the formula:

$$A_{d-d} \sim g_n g_e \mu_n \mu_e / r^3, \qquad (9)$$

where $g_n$ = 2.2632 is nuclear g-factor phosphorus, $g_e$ = 2.004 is electronic g-factor.

Consequently, cerium ions in state of $Ce^{3+}$ were associated with $PO_4$ and OH groups.

Table 6. Distance between Ce – P

| B, mT | $r_{experiment}$ (Å) | $r_{theory}$ (Å) | A (MHz) |
|---|---|---|---|
| 352 | 3.9±0.4 | 3.2 (Ca1 – P) | 0.47 |
| 304 | 3.4±0.3 | 3.4 (Ca2 – P) | 0.66 |

**EDNMR**

To obtain additional information about the nuclear environment, the EDNMR method was used, which is an analogue of the ENDOR method, but instead of the second radio frequency source, an additional microwave module with a sweep of 9.3 to 9.7 GHz is used. This method is therefore based on the phenomenon of double electron-electron resonance (DEER).

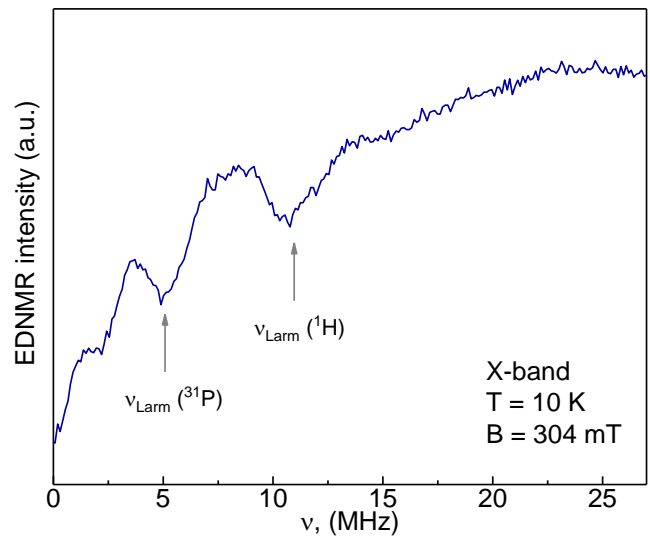

Fig. 7 EDNMR experiment for Ce-HAp recorded at T = 10 K

The EDNMR spectrum of the Ce-HAp sample recorded at T = 10 K is shown in Fig. 7. The observed signals with a frequency of ≈ 5 MHz from phosphorus nuclei and in the frequency range of 11-12 MHz from hydrogen confirm the presence of Ce ions in the Hap crystal structure, and not on the surface of the nanoparticles. Therefore, $Ce^{3+}$ is bound to OH groups and to phosphate groups.

**X-Ray irradiated samples**

The use of nitrate salts as initial reagents during the chemical synthesis of HAp leads to the incorporation of nitrate complexes $NO_3^-$ in the structure of the sample. In order to analyze the local environment, a nitrogen radical is used as a spin probe. Similar studies of irradiated calcium phosphates have been previously conducted in which radiation-induced paramagnetic centers located in hydroxyl or phosphate centers were identified [21]. Fig. 8 shows the EPR spectra of radiation-induced centers for Ce-HAp powders as-dried (green line) and the powder after heat treatment at 1300ºC in air (orange line). The interaction of the electron spin (S = ½) and the nuclear magnetic moment (I = 1, $^{14}N$) leads to the formation of three lines in Fig. 7 (orange line) with hyperfine splitting values $A_\perp$ = 6.87 mT and $A_\parallel$ = 13.2 mT.

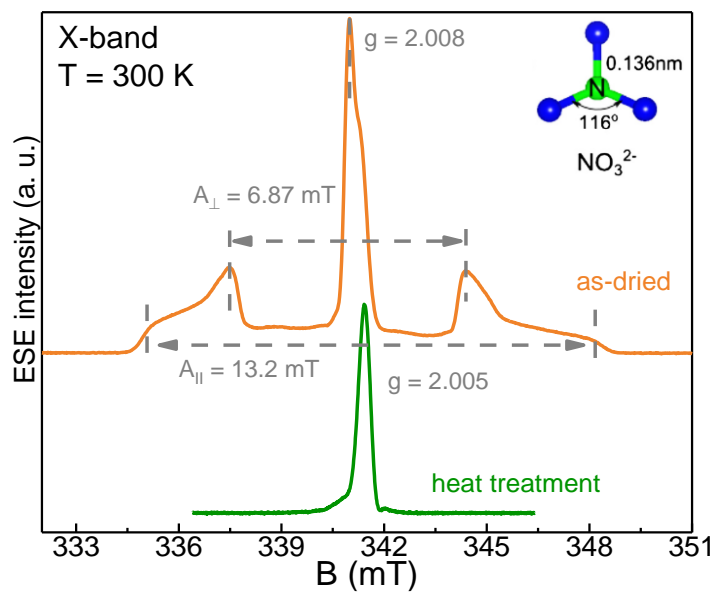

Fig. 8 ESE-EPR spectra recorded in the X-band at room temperature for samples after X-ray irradiation: Ce-HAp (as-dried, green line); Ce-HAp (after heat treatment at 1300ºC in air, orange line)

Heat treatment of the sample leads to the complete disappearance of synthesis by-products and a reduction in crystal lattice defects, as shown in Fig. 8 (green line). The EPR spectrum shows a small amount of carbonate radical ($CO_2^-$) with electron spin $S = 1/2$ and a resonance line at g-factor = 2.005. Carbonate anions act as spin traps and are able to capture electrons under the influence of X-rays, causing the impurities to become paramagnetic. Accordingly, the heat treatment procedure of the sample helps to reduce the amount of toxic nitrate radicals and increase the degree of crystallinity. Table 7 shows the dynamic characteristics of radiation-induced centers at room temperature.

Table 7. Distance between Ce – P

|  | $B$, mT | $T_1$, μs | $T_2$, μs |
|---|---|---|---|
| HAp-Ce (as-dried) | 340.9 | 22 | 2.0 |
|  | 337.4 | 23 | 2.3 |
| HAp-Ce (heat treatment) | 341.4 | 9.8 | 2.4 |

**Conclusion**

In this paper, we present a comprehensive analysis of synthesized hydroxyapatite powders doped with cerium ions. These powders were synthesized by chemical precipitation using calcium nitrate, cerium nitrate and ammonium hydrogen phosphate. XRD analysis showed that hydroxyapatite is the main phase in the synthesized powders. The EPR results show that during heat treatment by hot pressing at 1100 °C, Ce is included in the HAp structure in the trivalent state. For powders subjected to heat treatment in air at 1300 °C, Ce undergoes a partial transformation to the tetravalent state. In Ce-HAp (after drying), cerium is oxidized to the tetravalent state. The EPR spectrum recorded in the pulsed mode in the Kc range is due to contributions from the paramagnetic centers of cerium ions replacing calcium ions in two nonequivalent positions. ESEEM experiment, which establishes the location of cerium in the HAp crystal lattice and estimates the distance between calcium and cerium (approximately 3.5 Å). The results of the EDNMR method show that the cerium ions are bound to the OH group. However, hydrogen is not a structural unit of HAp. It can be assumed that hydrogen acts as a charge compensator. The results of powder studies after X-ray irradiation showed that the thermal treatment procedure helps to reduce the amount of toxic nitrate radicals and increase the degree of crystallinity.

**Funding:** This work was supported by the Russian Science Foundation (project no. 24-72-00161).

**Data Availability Statement:** Data can be available upon request from the authors.

**Conflicts of Interest:** The authors declare no conflicts of interest.